\documentclass[pra, preprint, superscriptaddress, nofootinbib, showpacs]{revtex4}
\usepackage[english]{babel}
\usepackage{graphicx}
\usepackage{dcolumn}
\usepackage{bm}
\usepackage{amsmath}
\usepackage{amsfonts, amssymb}
\usepackage{dsfont}
\usepackage{graphicx}
\usepackage{subfigure}
\usepackage{xcolor}
\usepackage{subfigure}
\def\pa{\partial}
\def\fr{\frac}
\def\Y{\Psi}
\def\nn{\nonumber}

\begin{document}

\title{Time-dependent configuration-interaction-singles calculation of the $5p$-subshell two-photon ionization cross section in xenon}

\author{Antonia Karamatskou}
\email{antonia.karamatskou@cfel.de}
\affiliation{The Hamburg Centre for Ultrafast Imaging, Luruper Chaussee 149, DE-22761 Hamburg, Germany}
\affiliation{Center for Free-Electron Laser Science, DESY, Notkestra\ss e 85, DE-22607 Hamburg, Germany}
\affiliation{Department of Physics, Universit\"at Hamburg, Jungiusstra\ss e 9, DE-20355 Hamburg, Germany}

\author{Robin Santra}
\email{robin.santra@cfel.de}
\affiliation{The Hamburg Centre for Ultrafast Imaging, Luruper Chaussee 149, DE-22761 Hamburg, Germany}
\affiliation{Center for Free-Electron Laser Science, DESY, Notkestra\ss e 85, DE-22607 Hamburg, Germany}
\affiliation{Department of Physics, Universit\"at Hamburg, Jungiusstra\ss e 9, DE-20355 Hamburg, Germany}

\begin{abstract}
The $5p$ two-photon ionization cross section of xenon in the photon-energy range below the one-photon ionization threshold  is calculated within the time-dependent configuration-interaction-singles (TDCIS) method. The TDCIS calculations are compared to random-phase-approximation (RPA) calculations [Wendin \textit{et al.}, J. Opt. Soc. Am. B \textbf{4}, 833 (1987)] and are found to reproduce the energy positions of the intermediate Rydberg states reasonably well. The effect of interchannel coupling is also investigated and found to change the cross section of the $5p$ shell only slightly compared to the intrachannel case. 
\end{abstract}


\maketitle

\section{\label{sec:level1}Introduction}
As a prime example of an atomic system exhibiting strong electron correlation effects, xenon and its $4d$ giant dipole resonance in the XUV one-photon absorption spectrum have been the topic of numerous studies over the past decades \cite{ederer64,Wen73,amu74,Sta82,Amu99,Mak09,Shi11}. The one-photon regime of the $4d$-subshell ionization was also studied within the time-dependent configuration-interaction-singles (TDCIS) approach~\cite{pab,krebs14,giant}.  

Nowadays, free-electron lasers such as FLASH~\cite{flashweb} or FERMI~\cite{fermiweb} provide highly intense coherent radiation in the XUV that allow for the investigation of matter beyond the linear-response regime~\cite{ackermann,WabnitzNAT02,allaria}. The availability of strong light pulses has increased the interest in nonlinear atom-light interactions and, in particular, the role of correlation effects in this regime. Recently, TDCIS was employed in the nonlinear regime to study XUV above-threshold ionization in xenon~\cite{Maz15}. The study revealed that two-photon ionization is a most sensitive tool to probe the effect of collective effects in the $4d$ shell. The electron correlation effects lead to two distinct resonance states constituting the giant dipole resonance~\cite{Maz15} which cannot be attributed to single particle-hole states~\cite{Che15}. 

Of course, multiphoton ionization in correlated atomic systems has been studied previously. Many-body perturbation theory was employed extensively over the last decades in order to investigate the multiphoton ionization of xenon~\cite{Jonsson:92,PhysRevA.36.5632}. The two-photon ionization of the $5p$ subshell in xenon was calculated within the random phase approximation (RPA) in various approximations and extensions~\cite{lhuillier87,Wendin:87,PhysRevA.36.4747}. 

In the present work we extend the investigation within TDCIS to the nonlinear ionization of the $5p$ shell of xenon and examine the quality of the intermediate states as calculated within the configuration-interaction-singles (CIS) scheme. In order to clarify the impact of interchannel coupling onto the $5p$ ionization process, interchannel and intrachannel results will be compared. While in the case of the $4d$ giant dipole resonance the intermediate states of the two-photon process are resonance states in the continuum, here the first photon excites an electron of the $5p$ shell to a Rydberg state, where it is ionized by the second photon. We will show that within the CIS approach the two-photon processes and the intermediate states involved are captured quite well. In comparing the results the RPA two-photon cross section results shall serve as a benchmark for the TDCIS calculations in order to establish a link between the two different approaches. 
Since TDCIS is a nonperturbative wave-function approach no diagrams or summation over intermediate states are involved. The aim is to investigate the quality of the TDCIS two-photon cross section. Moreover, the TDCIS one-photon cross section will be employed to demonstrate that the energy positions of the Rydberg states found in the one-photon excitation process and the two-photon ionization process are consistent, and that the ratio between the cross sections at photon energies of consecutive Rydberg states is different in the one- and the two-photon processes. 

\section{\label{sec:2}Theory and Method}
The time-dependent Schr\"odinger equation of the full $N$-electron system reads
\begin{equation}
 i \fr{\pa}{\pa t} |\Y^N(t)\rangle = \hat {H} (t) |\Y^N(t)\rangle . \label{schr}
\end{equation} 
The $N$-electron wave function is expanded in the one-particle--one-hole basis~\cite{pab}:
\begin{equation}
|\Psi^N (t) \rangle=\alpha_0(t) |\Phi_0 \rangle+ \sum_{i,a}\alpha_i^a(t)|\Phi_i^a \rangle,
\end{equation}
where the index $i$ and $a$ denote an initially occupied and a virtual orbital, respectively. The indices of electron and hole, respectively, represent the full set of quantum numbers $n,l,m$. The total spin of the system is set to $0$, since only spin singlets are considered. Here, $|\Phi_0 \rangle$ symbolizes the Hartree-Fock ground state. The full time-dependent Hamiltonian has the form
\begin{align}
 \hat{H}(t)&= \hat{H}_0+\hat{H}_1+\hat{\mathbf{p}} \cdot \mathbf{A}(t)\\
&=\sum_{n=1}^N \left(\fr{\hat{\mathbf{p}}_n^2}{2}-\fr{Z}{|\hat{\mathbf{r}}_n|} \right) +\hat{V}_{\mathrm{ MF}}-E_{\mathrm{HF}}\nn\\
&+ \fr{1}{2} \sum_{n\neq n'}^N \fr{1}{|\hat{\mathbf{r}}_n-\hat{\mathbf{r}}_{n'}|}-\hat{V}_{\mathrm{ MF}}\nn\\
&+\sum_n \hat{\mathbf{p}}_n \cdot \mathbf{A}(t), \label{ham}
\end{align}
where the kinetic energy $\hat{T}=\sum_{n=1}^N \hat{\mathbf{p}}_n^2/2$, the nuclear potential $\hat{V}_{\rm nuc}=-\sum_{n=1}^NZ/|\hat{\mathbf{r}}_n|$, the mean-field potential $\hat{V}_{\rm MF}$ and the Hartree-Fock energy $E_{\mathrm{HF}}$ constitute the one-body operator $\hat{H}_0$, $\hat{H}_1=\fr{1}{2} \sum_{n\neq n'}^N \fr{1}{|\hat{\mathbf{r}}_n-\hat{\mathbf{r}}_{n'}|}-\hat{V}_{\rm MF}$ describes the Coulomb interactions beyond the mean-field level, $\hat{\mathbf{p}}\cdot\mathbf{A}(t)$ is the light-matter interaction in the velocity form in the dipole approximation (assuming linear polarization), and $\mathbf{A}(t)$ is the vector potential of the electromagnetic field (here a vacuum ultraviolet pulse) that interacts with the electronic system. The Schr\"odinger equation is solved numerically via time propagation \cite{XCIDprog} as described in Ref.~\cite{pab}.

Being a multichannel approach, TDCIS allows for the disentanglement of all channels $i$ open to ionization. Furthermore, owing to the dipole selection rules one-photon excitation or two-photon ionization populations can be discerned in the perturbative limit through the different, yet uniquely determined, angular momenta of the final states. For this purpose the matrix elements of the ion density matrix are calculated~\cite{pab} and the populations of the cationic states are considered
\begin{eqnarray}
\rho_{ii}^{\mathrm{IDM}}&=&\mathrm{Tr}_a \left(|\Psi^N(t)\rangle\langle\Y^N(t)|\right)_{ii}\\
&=&\sum_{l_a}\sum_{n_a,m_a} \langle\Phi_i^a   |\Y^N(t)  \rangle \langle \Y^N(t)| \Phi_i^a  \rangle,\label{second}
\end{eqnarray}
where the trace extends over all virtual indices $n_a,l_a,m_a$. 
Here, the quantities $\sum_{n_a,m_a} \langle\Phi_i^a   |\Y^N(t)  \rangle \langle \Y^N(t)| \Phi_i^a  \rangle$ 
are of special interest because, thereby, the depopulations of a channel $i$ can be classified according to the 
angular momenta of the virtual orbital $l_a$. This means that the depopulations leading to a particular angular momentum 
of the corresponding final state can be identified properly, as long as only one- and two-photon processes play a role.
 In the case considered here, the channel of interest is the $5p$ ionization. 
 After absorbing one photon the $5p$ electron is promoted to a final state 
 with angular momentum $l=0$ or $l=2$, while the absorption of two photons 
 of a $5p$-shell electron leads to a final state with $l=1$ or $l=3$. Since these 
 angular momenta are disjunct for the two different processes, the one- and two-photon 
 ionization probabilities can be calculated by distinguishing the corresponding $5p$-shell 
 depopulations according to the angular momenta of all possible final states.

The cross sections $\sigma^{(N)}$ are then obtained as the quotient of the corresponding $5p$ depopulation by $N$ photons, denoted as $P_N$, and the fluence available for the process under investigation $F^{(N)}=\int dt [I(t)/\omega_{\mathrm{ph}}]^N$, where $I(t)$ is the intensity envelope:
\begin{equation}
\sigma^{(N)}=\frac{P_N}{F^{(N)}},
\end{equation}
 as illustrated in greater detail, for instance, in Refs.~\cite{Til15,AK17}. Of course, the pulse employed for the calculations is sufficiently weak to satisfy the conditions of the perturbative regime and to guarantee that multiphoton ionization beyond two-photon ionization remains negligible.

\section{\label{sec:3}Results and Discussion}

\begin{figure}[!tb]
 \includegraphics[width=\linewidth]{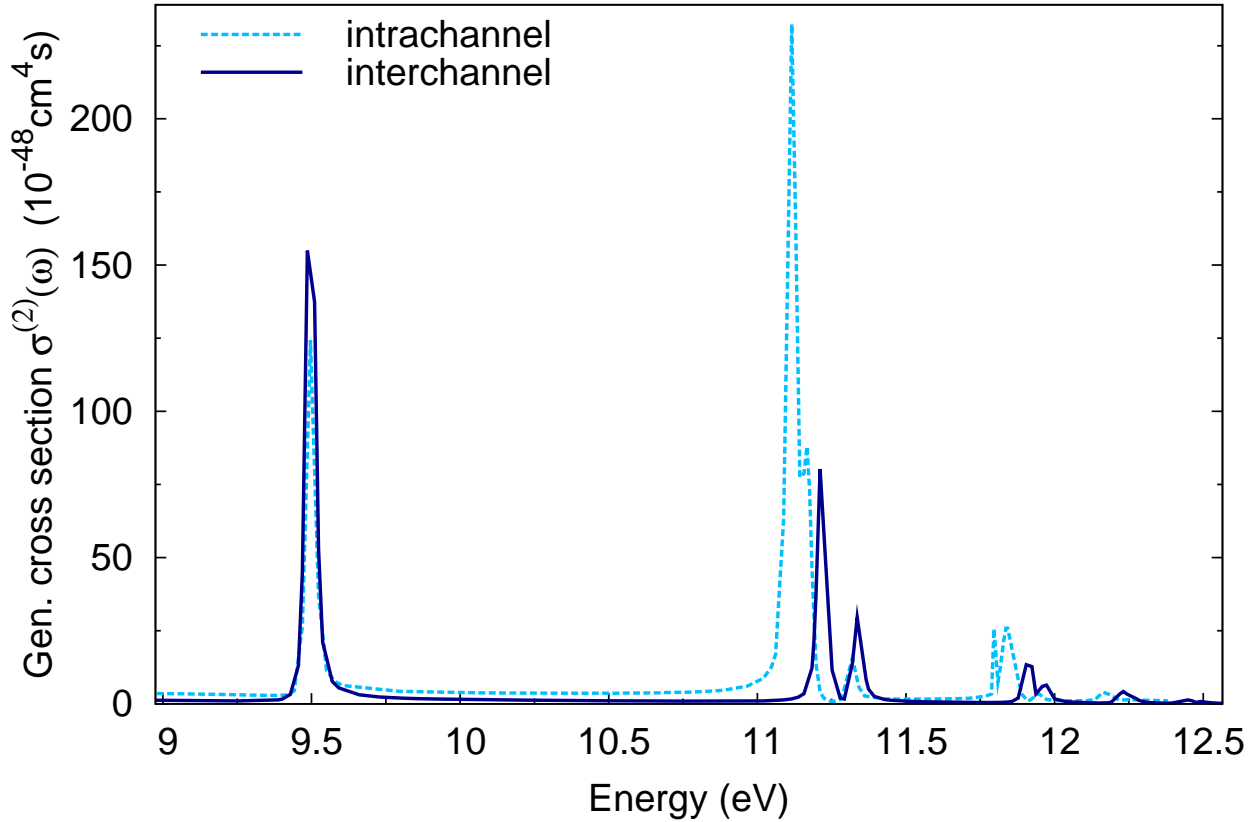}
 \caption{Xenon $5p$ two-photon ionization cross section in the interchannel and intrachannel coupling schemes below the one-photon ionization threshold. The series of peaks reflects the energy positions of the first Rydberg states associated with an excitation from the $5p$ shell.}
 \label{2phintercomp}
\end{figure}

The two-photon generalized cross sections are calculated for each photon energy in the interchannel and intrachannel coupling model using a Gaussian pulse with a pulse duration of $3000$~a.u., such that the energy width is small and, hence, can span maximally one intermediate resonance state. In that way the pulse mimics a continuous-wave light field. The results for the two-photon ionization cross section are compared in Fig.~\ref{2phintercomp}. The peaks in both spectra can be attributed to the first several $5p$ Rydberg states, being $5p^{-1}6s$, $5d$, $7s$, $6d$, $8s$, and so forth. They participate as intermediate states in the two-photon ionization process. The $ns$ peaks are not altered in the interchannel scheme and can be nicely compared to the intrachannel case, whereas for the $nd$ lines a splitting is observed in the intrachannel case. Also, the two schemes differ in the ratio of the line strengths of consecutive peaks. Note that the line widths and peak strengths in Fig.~\ref{2phintercomp} are limited by the Gaussian pulse duration used.
 
 \begin{figure}[!tb]
 \includegraphics[width=\linewidth]{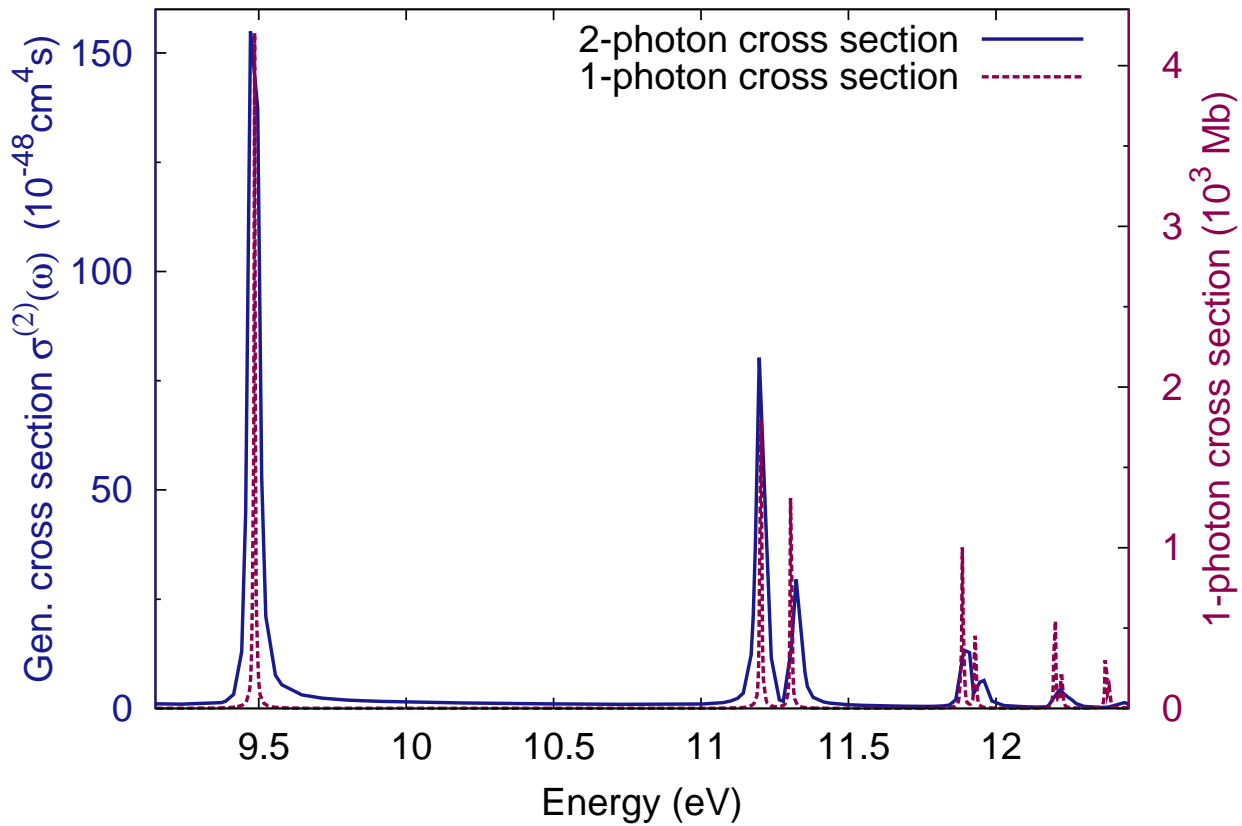}
 \caption{One-photon excitation and two-photon ionization cross sections of the xenon $5p$ shell. The energy positions of the Rydberg states, which are intermediate states in the two-photon process, are reflected in both processes and coincide nicely.}
 \label{1ph2ph}
\end{figure}
 
In order to legitimate the quality of the positions of the Rydberg peaks visible in the interchannel two-photon ionization cross section we consider also the one-photon process in this photon-energy regime where the $5p$ electron is excited into Rydberg states. The one-photon excitation cross section is calculated employing the autocorrelation-function technique described in Ref.~\cite{krebs14}. For the calculation of the autocorrelation function a total propagation time of $5\times 10^4$~a.u. is used which yields narrow cross-section peaks. 

In Fig.~\ref{1ph2ph} both cross sections are shown together, i.e., the interchannel two-photon ionization cross section shown in Fig.~\ref{2phintercomp} is compared to the one-photon excitation cross section calculated using the autocorrelation function. The energy positions of the Rydberg states that result both in the one-photon excitation and the two-photon ionization cases coincide nicely. There is merely a slight shift of the two-photon cross section to higher energies starting from the third peak which could be due to a combination of the pulse duration used in the calculations and the choice of the photon energy points, because every single cross section point is calculated in a full TDCIS propagation using this very photon energy. 

However, there is a significant difference in the relative heights of the peaks between the one-photon excitation and the two-photon ionization processes. The two-photon ionization peaks decrease more rapidly in height than the one-photon excitation peaks, and also the ratio of the heights in the $nd$-$(n+2)s$ groups is different. This suggests that the intermediate states acquire a different weight in the two-photon transition matrix element. The phenomenon of a different weight of intermediate resonances was also observed in previous work on $4d$ two-photon (above-threshold) ionization of xenon where the intermediate resonance states constitute the giant dipole resonance~\cite{Maz15,Che15}. In that case, however, interchannel coupling was identified to play an important role and to lead to intermediate states that cannot be attributed to single particle-hole states. The intermediate states in the $4d$ two-photon above-threshold ionization process were found to interfere and to lead to a substantially broadened two-photon cross section. In the case of the $5p$ ionization the intermediate states correspond to long-lived Rydberg states that do not interfere with each other.

Regarding the absolute strengths of the spectral lines we note that TDCIS might deviate from the experimental values, because the quality of the matrix elements within the CIS approach is restricted. However, qualitatively, the TDCIS two-photon ionization cross section can describe the positions of the intermediate states. It would be interesting to compare experimental two-photon cross-section measurements to the results presented in this work, similar to the comparison performed in Ref.~\cite{Maz15}, in order to estimate the importance of deviations from the CIS space in the two-photon ionization process.

\begin{figure}[!tb]
 \includegraphics[width=\linewidth]{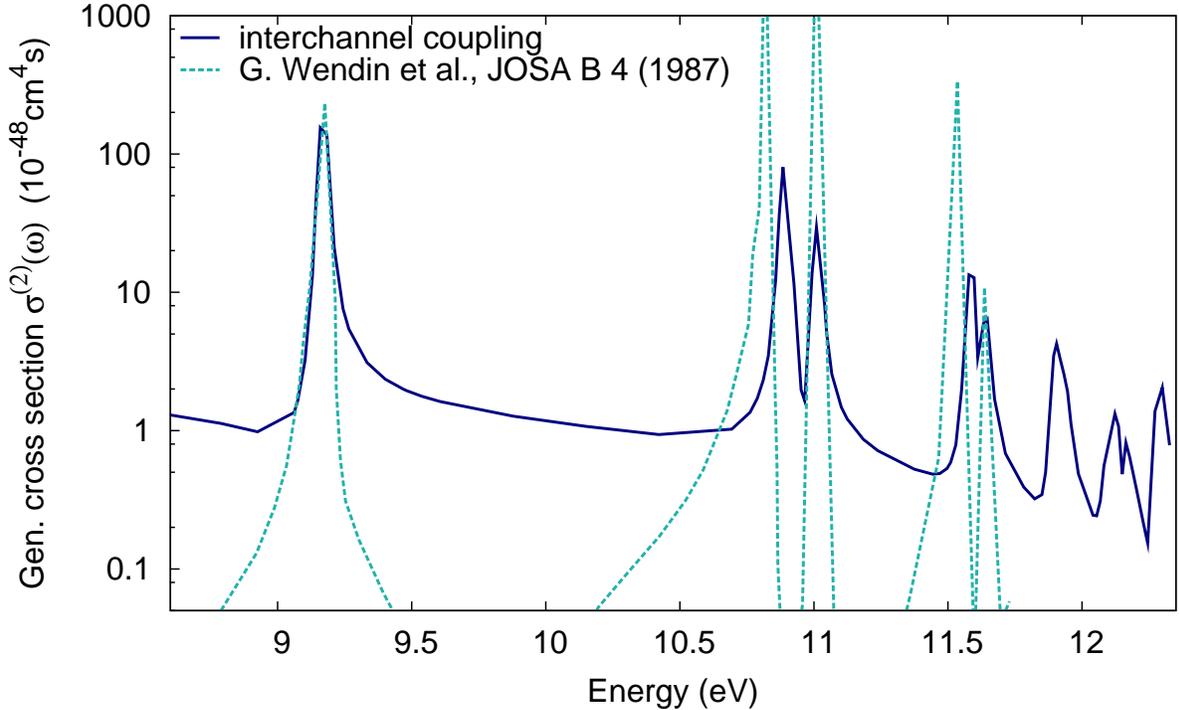}
 \caption{Comparison of the TDCIS two-photon ionization cross section with the RPAE results of Ref.~\cite{Wendin:87}. The agreement in the positions of the intermediate states is reasonable, even though the TDCIS positions are slightly shifted to higher energies. The heights, however, are underestimated by TDCIS.}
 \label{internist}
\end{figure}

Finally, we compare our results for the two-photon cross section with the results presented in Ref.~\cite{Wendin:87}. First, we note that in strong contrast to the results of the local-density approximation (LDA) and the local-density random-phase approximation (LDRPA) presented in Ref.~\cite{Wendin:87} TDCIS produces no artificial structures in the two-photon cross section. Bound resonance states as well as resonance states in the continuum appear quite naturally within the TDCIS two-photon cross sections if they are accessible by one photon and, thus, can be identified immediately as intermediate states.

Therefore, we compare TDCIS to the more elaborate model of the RPA-with-exchange (RPAE) approach. When shifting the $5p$ binding energy about $0.3$~eV down to the energy found within RPAE the cross-section curves for the two-photon cross section agree nicely as far as the positions of the Rydberg states are concerned. This is shown in Fig.~\ref{internist} for the range available for comparison as provided in Ref.~\cite{Wendin:87}. There is a slight shift to higher energies visible in the second and fourth Rydberg peak. The fact that the absolute heights of the peaks corresponding to the Rydberg intermediate states differ in the two methods can be attributed to the pulse duration used in the TDCIS calculations, which affects the widths and the heights of the peaks. However, the same number of peaks is observed in both approaches and also their spacing agrees reasonably well. This suggests that TDCIS results can be meaningfully compared to many-body perturbation theory calculations such as RPAE including screening and double excitations.

\section{\label{sec:4}Conclusion}
The presented results and, in particular, the comparison with RPAE calculations support the assumption that TDCIS is an adequate tool to calculate two-photon generalized cross sections that involve real or virtual intermediate states. In the case of the $5p$ two-photon ionization process, the intermediate states are Rydberg states that manifest themselves as peaks in the one-photon excitation cross section. Their positions are well reproduced in the two-photon ionization cross section, but the relative heights of consecutive peaks can differ between the one-photon excitation and the two-photon ionization processes. 

Interchannel coupling is not found to significantly alter the $5p$ two-photon ionization cross section compared to the intrachannel case. This stands in strong contrast to the two-photon above-threshold ionization of the $4d$ subshell. 

\section{Acknowledgments}
We would like to thank G\"oran Wendin for inspiring discussions. AK is funded by The Hamburg Centre for Ultrafast Imaging through the Louise-Johnson Fellowship.

\bibliographystyle{apsrev}
\bibliography{bibl.bib}

\end{document}